\documentstyle[11pt,newpasp,twoside,epsf]{article}
\def\edcomment#1{\iffalse\marginpar{\raggedright\sl#1\/}\else\relax\fi}
\reversemarginpar

\begin{document}
\title{GAIA survey of galactic eclipsing binaries}
\author{Toma\v{z} Zwitter}
\affil{University of Ljubljana, Department of Physics, Jadranska 19,
1000 Ljubljana, Slovenia, tomaz.zwitter@uni-lj.si}

\begin{abstract}
General importance and capabilities of observations of eclipsing binaries 
by the forthcoming ESA mission GAIA are discussed. Availability of 
spectroscopic observations and a large number of photometric bands on board 
will make it possible to reliably determine physical parameters for 
$\sim 10^5$ binary stars. It is stressed that current methods of object 
by object analysis will have to be modified and included in an automatic 
analysis pipeline. 
\end{abstract}

\section{Introduction}

GAIA is the approved Cornerstone 6 mission of the European Space Agency. 
Its main goal is to observe up to a billion stars in our Galaxy and 
obtain their astrometric positions on a micro-arc sec level, multi-band 
photometry in 15 different optical and near-IR bands, as well as 
spectroscopic observations within the  250~\AA\ interval around the Ca~II IR 
triplet. Perryman et al. (2001), ESA-SCI(2000)4 and Munari (1999, 2001) 
are useful introductions to the general properties of the mission, 
its overall astrophysical importance and diagnostic capabilities of  
the spectral window chosen for the spectrograph. Munari (2002) 
discusses the potential of GAIA to observe peculiar stars. Here 
we focus on galactic eclipsing binaries which are rapidly becoming one of the 
areas where the harvest of GAIA's results will be the richest. 

In the next section we discuss general importance and capabilities of 
observations of eclipsing binaries by GAIA. Next the mode in which 
spectroscopic observations are obtained is briefly discussed, together with 
sources of noise, intensity of the background and 
the possibility for spectral tracing overlaps.
All this leads to the estimates on expected radial velocity accuracy as a 
function of magnitude and spectral type of the target. Finally we present 
results from observations of real binary stars in the GAIA-like mode and 
so demonstrate GAIA's accuracy in determination of basic stellar parameters. 
We conclude with some remarks on automatic reduction of the gigantic 
database of GAIA's observations of galactic eclipsing binaries. 
  
\section{Importance and capabilities of binary star observations}

We shall illustrate the capabilities of observations of binary stars 
for two cases: stars brighter than $V=15$, and stars brighter than $V=17$.
For the former class the errors on single-star parameters obtained from a 
combination of astrometry, photometry and spectroscopy will be less than
50~K for the effective temperature, less than 0.1 for gravity, less than 0.2 for 
metallicity and less than 10\%\ for distance (for distances within 10~kpc)
(see ESA-SCI(2000)4).
These values could be checked by observations of some $4\times 10^5$ 
eclipsing binaries with $V<15$. Of these some $\sim 10^5$ will be 
double-lined. It is not too optimistic to expect that one can obtain a 
few ten-thousand excellent orbital solutions with secure determination of 
stellar parameters, a huge increase over the current number of 
reliably solved eclipsing binaries. Apart from 
measurement of stellar parameters that cannot be independently determined 
for single stars (mass and to some extent radius) GAIA's orbital solutions 
of bright eclipsing binaries will have the potential to determine their 
distances even for objects much farther than 10~kpc, the base-lined 10\%\ 
accuracy limit for astrometry. 

\begin{figure}
\plotone{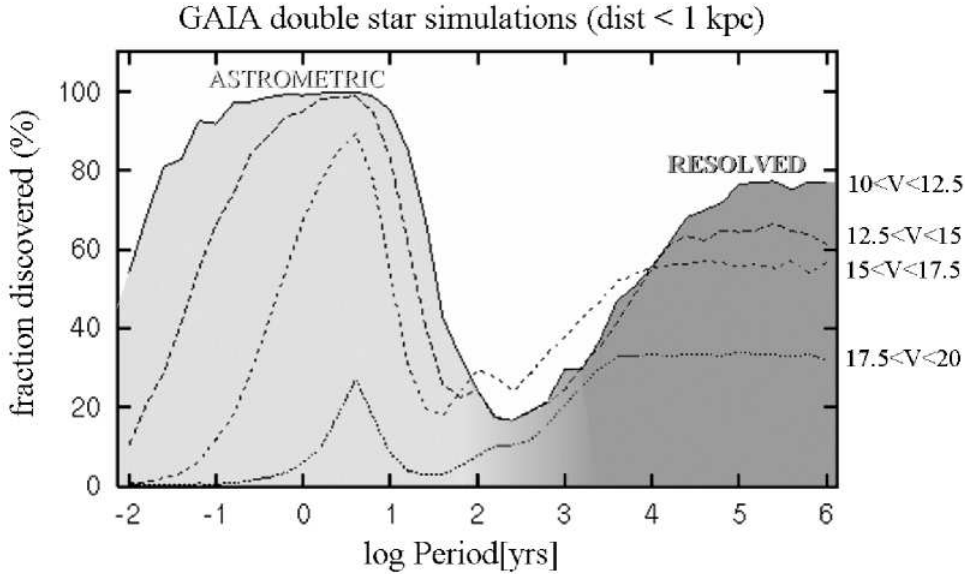}
\caption{Fraction of astrometric and resolved binary stars discovered by GAIA 
as a function of orbital period and for different magnitude ranges.
Simulations were done for a distance 
limited sample ($d<1 $~kpc). Adapted from ESA-SCI(2000)4. 
}
\end{figure}

For fainter targets ($V<17$, distance $< 10$~kpc) GAIA will provide observations
of $\sim 5\times 10^7$ stars. It is important to note that all these stars 
will be observed in exactly the same way. Some of the stars will be single, but 
others will be binaries with orbital periods from hours to hundreds of years. 
The obvious question one can address with this database is how binarity 
influences stellar evolution.
Eclipsing binaries will be observed and discovered also at fainter magnitudes.
Good quality light curves could be obtained but spectroscopic observations will 
be of limited accuracy. Current galactic models (see ESA-SCI(2000)4) 
predict discovery of $\sim 7 \times 10^6$ eclipsing binaries at $V< 20$. 

GAIA will survey a wide range of binary star separations and orbital periods 
(Fig.~1). At the longest orbital periods ($>100$~yrs) the components will be 
resolved on the sky and their binary nature could be judged from their equal 
parallax, proper motion and a similar radial velocity. Closer binaries, down 
to orbital periods of days, will be discovered by the curved, 
non-linear proper motion path. Finally, spectroscopy will be useful to 
discover even shorter period binaries, not plotted in Fig.~1.  
As seen from the Figure the fraction of 
discovered binaries is significant, thus GAIA will be able to study the 
binarity phenomenon for the whole range of orbital periods.

\section{Performance of the spectrograph}

The spectrograph aboard GAIA will have a square primary mirror with the surface 
area of $0.5\times0.5$~m$^2$ and the field of view of $1.65\deg \times 1.6\deg$. 
It is a slitless spectrograph with stars drifting across the focal plane at the 
rate of 60"/s. This gives a fixed exposure time of 99 seconds for each transit 
of the star across the focal plane. Because of a precessional motion of the 
spin axis of the satellite the whole celestial sphere will be covered, and 
each position on the sky will be observed on average 100-times during a 5-year 
mission. 

The spectral interval is base-lined to 8490-8750~\AA. It includes Paschen lines 
P13-P16, Ca~II IR triplet as well as numerous N~I, Fe~I, Ti~I, Cr~I, Si~I, 
Ni~I, S~II, Mg~I and Mn~I lines (Munari 2001). The resolution will be most 
likely between 10000 and 15000. 

The performance of the spectrograph will be limited by shot noise, RON, zodiacal
background and - to some extent - spectral tracing overlaps. Zodiacal background 
was discussed by Zwitter (2002). Overlapped spectral tracings will not 
present a significant problem in terms of spectral line blending, as stars' 
angular separation on the sky guarantees that lines generally do not overlap. 
Moreover the scanning law is such that two spectral tracings overlapping on 
one transit will be recorded separately on the next occasion of observation of 
the same field. The main contribution of spectral overlaps will thus be an 
increase in the effective background. This will become significant for stars
fainter than $V\sim 17$ (except in the galactic plane where stellar overlaps 
will be more common). 

Figure 2 summarizes the expected radial velocity accuracy 
as a function of magnitude and spectral type. The zodiacal background of 
$V=21.5/$arcsec$^2$ was assumed, which is typical for an object on the 
ecliptic plane. This figure is the same as Fig.~4 of Zwitter (2002), but 
calculated for updated parameters of the spectrograph mentioned above. 
At the bright magnitude end higher resolution always gives better accuracy. 
For the faintest targets the accuracy achieved by the highest resolution 
(0.25~\AA/pix) is degraded due to the fact that the Zodiacal background
per wavelength bin does not depend on resolution, while the stellar 
signal is more dispersed at higher resolutions. Similar results were 
obtained also by observations of real stars (Munari et al.\ 2001b). 

\begin{figure}
\plotone{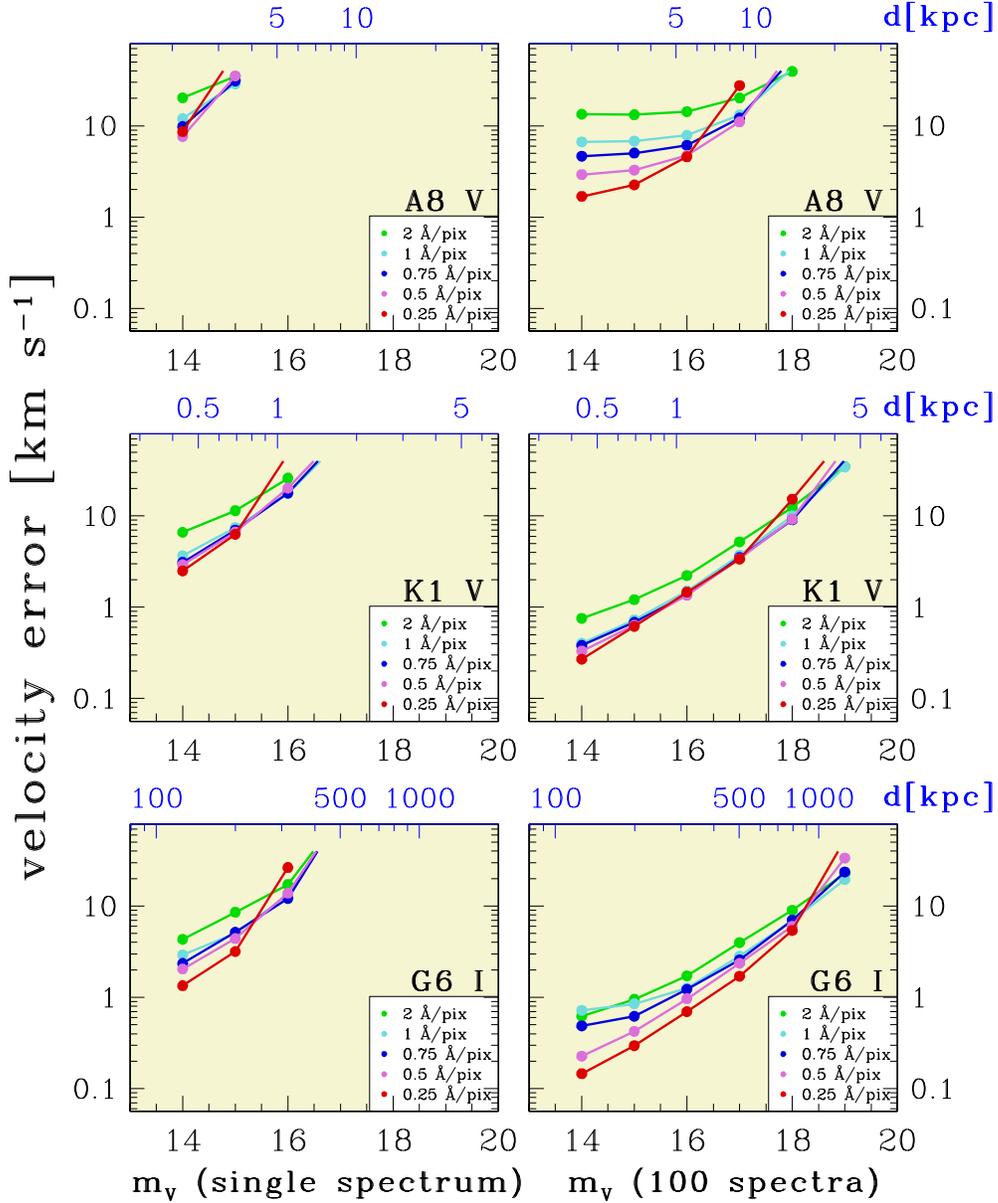} 
%/home/tomaz/dos/RVIII/mytalk/june02withrotvel_allplot74.eps
%\plotone{ZwFig2bw.eps}
% /home/tomaz/hipparcos/sep01/june02mismatch/withrotresults/june02withrotvel_allplot74bw.eps
\caption{Average radial velocity determination error as a function of stellar
magnitude, spectral type and the used spectral dispersion (assuming 2-pixels 
per resolution element). Top labels at each graph quote a representative 
distance if no interstellar absorption is present. Left column is 
for single transit spectra, while the right is for a mission-averaged 
spectrum (100 transits).}
\end{figure}

\begin{figure}
\plotfiddle{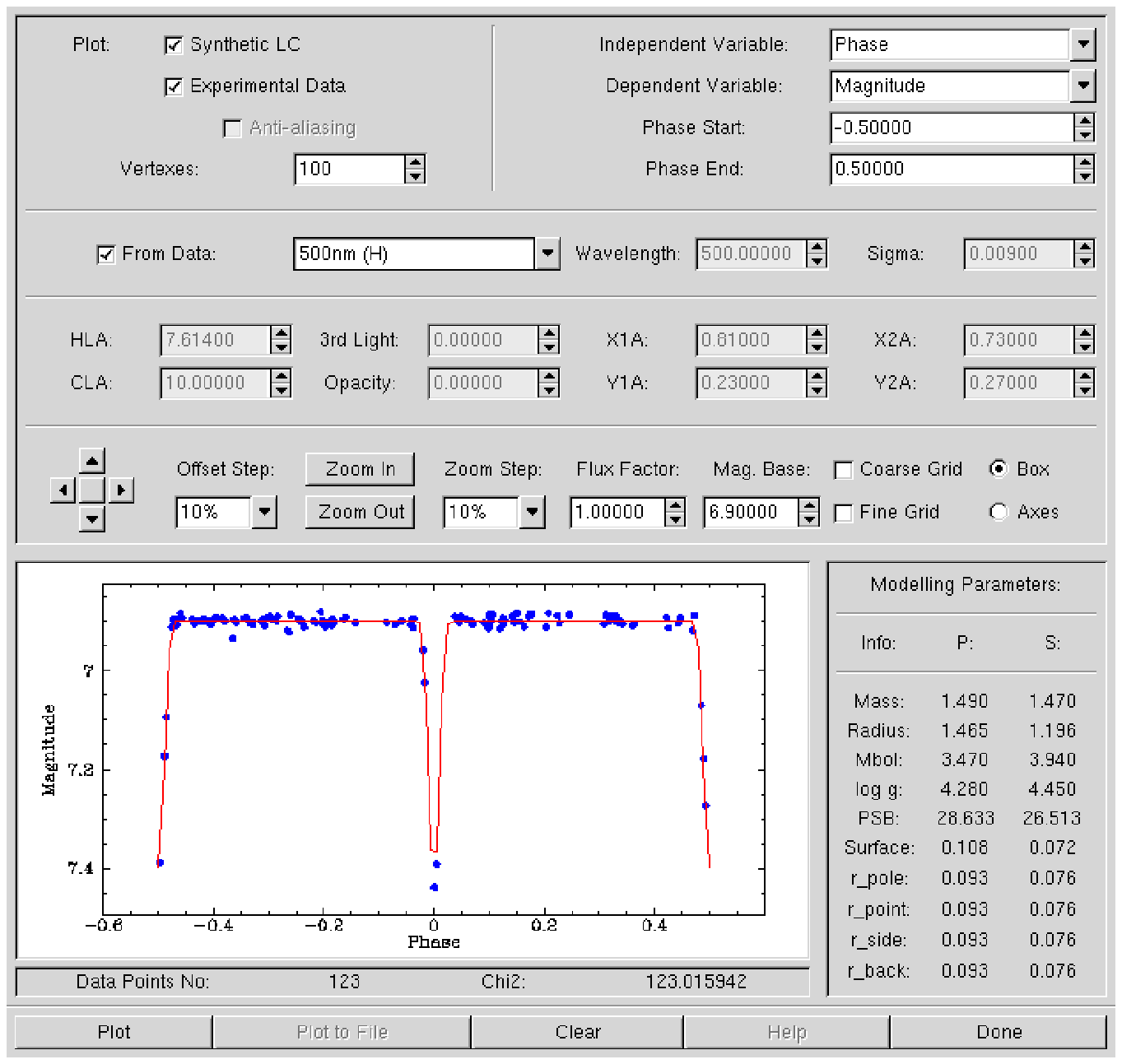}{3.3in}{0}{65}{65}{-125}{0}
%\plotone{ZwFig3.eps} 
\caption{Graphical user interface to run the Wilson-Devinney program
(A.\ Pr\v{s}a 2002, private communication).}
\end{figure}

\section{Observations and analysis of galactic eclipsing binaries}

Most binaries observed by GAIA will contain dwarfs of type G or K. 
Radial velocity accuracies mentioned above demonstrate that the GAIA's 
spectrograph will be able to secure useful radial velocity curves for 
binaries down to $V\sim 17$. GAIA will also obtain $\sim 100$ photometric 
observations of each binary in each of its 11 narrow- and 4 broad-band 
photometric filters. These can be used to determine stellar temperatures, 
gravities as well as provide some information on the metallicity. Typical 
errors for single stars will be 125~K in $T_{eff}$, 0.25 in $\log g$ and 
0.4 in $[Z/Z_o]$ (ESA-SCI(2000)4). These values can be further improved by 
spectra fitting. 

For eclipsing binaries the light curve and spectra analysis can improve 
the accuracy further. Munari et al.\ (2001a) published spectroscopic 
observations in the GAIA spectral window of three eclipsing binaries 
discovered by the Hipparcos satellite. Only three Hipparcos photometric bands 
with very limited S/N instead of the 14 bands previewed for GAIA were used in 
light curve analysis. Still the results have very acceptable 
error-bars: $\sim 2$~\%\ in masses, $\sim 100 $~K in temperatures, 
$\sim 0.1$ in $\log g$, and $\sim 7$\%\ in distance. 
These results are encouraging and demonstrate that GAIA data will be suitable 
to obtain accurate binary solutions. Reliable information on temperature, 
supplemented by actual fitting of the template spectra at each of the 
$\sim 100$ transits, opens the possibility to detect and analyze magnetic  
spots on the surfaces of eclipsing binaries which should be quite common 
for cool dwarfs. Such variability was also detected in yet unpublished 
eclipsing binaries discovered by Hipparcos that we have been observing in 
Asiago in the past 3 years. 

As mentioned above the number of eclipsing binaries discovered 
by GAIA will reach hundred-thousands. This means that the analysis 
will have to be fully automatized. All steps that were so far performed 
by astronomers, like the choice of initial fitting values, will have to be 
incorporated in an integrated reduction pipeline. One of the possibilities for 
the initial value determination is the minimum distance method. Wyithe and 
Wilson (2001, 2002) used an automatic reduction technique to analyze the 
OGLE photometric database. As expected they concluded that the sum of 
radii of both stars and the ratio of surface brightnesses can be reliably 
determined, while the ratios of radii, ratios of luminosities, inclination and 
eccentricity remain largely unconstrained by the OGLE photometric light curves. 
It should be stressed that the presence of spectroscopy aboard GAIA will 
largely remove these problems. A program running the 
Wilson-Devinney code through a suitable graphical interface (Fig.~3)
has been developed as a first step toward automatizing of light 
curve and spectral analysis. 

\section{Conclusions}

The GAIA mission has the potential to discover and characterize thousands 
of new peculiar and binary stars that could be followed up by additional 
observations by automatic ground based telescopes. It is crucial that 
suitable codes are developed for a correct recognition of eclipsing and 
non-eclipsing binaries on a central reduction pipeline. Once recognized 
the light curves and spectra will need to be analyzed with an automatic 
and reliable automatic procedures. 

The unique potential of GAIA among the forthcoming space missions is 
the availability of spectroscopy on board and an unprecedented number of 
targets that are to be observed. The size of the database at the end of the
mission will exceed 10 TB. However, as correctly pointed out by 
Peter Eggleton, more data can mean less understanding. To avoid this it 
is very important to think now of the new era when even peculiar binaries 
will be observed in large numbers. 
\medskip

{\bf Acknowledgment.} This work has been supported by a grant from the 
Slovenian Ministry of Education, Science and Sport.

\end{document}